\documentclass[runningheads]{llncs}
\usepackage[T1]{fontenc}
\usepackage{graphicx}
\usepackage[linesnumbered,ruled,vlined]{algorithm2e}
\usepackage{mdframed}
\usepackage{booktabs}
\usepackage{xcolor}
\usepackage{multirow}
\usepackage{float}
\usepackage{subcaption}
\usepackage{array}
\usepackage{soul}
\usepackage[most]{tcolorbox}  
\usepackage{hyperref}
\newtcolorbox[auto counter, number within=section]{promptbox}[2][]{%
  float*=tbp, 
  colback=white,
  colframe=black,
  title={Prompt~\thetcolorbox: #1},
  label={#1}
}
\begin{document}
\title{In Numeris Veritas: An Empirical Measurement of Wi-Fi Integration in Industry}
%
%
\author{Vyron Kampourakis\inst{1}\orcidID{0000-0003-4492-5104} \and
Christos Smiliotopoulos\inst{2}\orcidID{0000-0001-7530-7152} \and
Vasileios Gkioulos\inst{1}\orcidID{0000-0001-7304-3835} \and
Sokratis Katsikas\inst{1}\orcidID{0000-0003-2966-9683}}
\authorrunning{V. Kampourakis et al.}
%
\institute{Norwegian University of Science and Technology, 2802 Gjøvik, Norway \email{\{vyron.kampourakis, vasileios.gkioulos, sokratis.katsikas\}@ntnu.no} \and  University of the Aegean, 83200 Karlovasi, Greece \email{csmiliotopoulos@aegean.gr}}
\maketitle              
\begin{abstract}

Traditional air gaps in industrial systems are disappearing as IT technologies permeate the OT domain, accelerating the integration of wireless solutions like Wi-Fi. Next-generation Wi-Fi standards (IEEE 802.11ax/be) meet performance demands for industrial use cases, yet their introduction raises significant security concerns. A critical knowledge gap exists regarding the empirical prevalence and security configuration of Wi-Fi in real-world industrial settings. This work addresses this by mining the global crowdsourced WiGLE database to provide a data-driven understanding. We create the first publicly available dataset of 1,087 high-confidence industrial Wi-Fi networks, examining key attributes such as SSID patterns, encryption methods, vendor types, and global distribution. Our findings reveal a growing adoption of Wi-Fi across industrial sectors but underscore alarming security deficiencies, including the continued use of weak or outdated security configurations that directly expose critical infrastructure. This research serves as a pivotal reference point, offering both a unique dataset and practical insights to guide future investigations into wireless security within industrial environments.

\end{abstract}

\section{Introduction}
\label{S:intro}

At the epicenter of Industry 4.0, industrial systems and Critical Infrastructures (CI) are rapidly integrating modern technologies, including the Internet of Things (IoT) and Industrial IoT (IIoT), Artificial Intelligence (AI) and Machine Learning (ML), and cloud computing and analytics. This convergence is progressively dissolving the traditional air gaps that once characterized Operational Technology (OT) environments. Wireless technologies are no exception, enabling increasingly flexible, agile, and cost-effective operational paradigms, crucial for supporting IIoT deployments and enhancing overall flexibility and efficiency~\cite{VK:2023}. The predominant wireless protocol for Wireless Local Area Networks (WLANs) is IEEE 802.11, universally known as Wi-Fi. The advent of Wi-Fi 6 (802.11ax) and the emerging Wi-Fi 7 (802.11be) has brought significant advancements in throughput, latency, security, and overall performance. These substantial improvements render next-generation Wi-Fi a viable and increasingly attractive solution for meeting the stringent reliability, low-latency, and high-density requirements of mission-critical industrial environments.

However, the integration of Wi-Fi into industrial environments introduces significant security challenges. Unlike formerly isolated Operational Technology (OT) ecosystems, wireless-enabled networks inadvertently extend network access beyond physical perimeters, as Wi-Fi signals can propagate outside facility walls. This expanded attack surface enables adversaries with commodity radio devices to conduct passive reconnaissance, such as wardriving, or to actively attempt unauthorized access from a distance, without ever stepping foot inside the premises~\cite{wiPeep}. A compelling real-world example occurred in 2022, where threat actors utilized drones equipped with Wi-Fi Pineapple modules to capture employee credentials; subsequent drone deployments actively attempted unauthorized access to a financial firm's internal Wi-Fi network by exploiting signal leakage beyond the building's physical boundaries~\cite{incident2022}. Another notable real-world example occurred ahead of Russia’s invasion of Ukraine, where Advanced Persistent Threat (APT) actors employed a so-called \textit{nearest neighbor} attack to covertly infiltrate a target organization’s internal Wi-Fi. Namely, they compromised adjacent organizations and hijacked dual-homed systems with active Wi-Fi radios, exploiting previously harvested credentials to access enterprise networks without ever being physically present~\cite{volexity2024}. These incidents powerfully underscore the central problem: as industrial operations become increasingly interconnected via Wi-Fi, are existing security practices adequately protecting these critical networks? Despite these evident risks, there remains a dearth of empirical data and research on the actual prevalence of Wi-Fi in industrial settings and the specific security configurations deployed in the field. 

To address this critical gap, this study sets out to investigate the following research question: What is the current state of Wi-Fi deployment and associated security configurations in real-world industrial environments? To do this, we adopt a data-driven approach, leveraging the global crowdsourced WiGLE database~\cite{wigle} as our lens into real-world Wi-Fi usage. The WiGLE platform aggregates publicly submitted, geotagged wireless network metadata, making it an invaluable resource for understanding real-world wireless infrastructure at scale. By systematically identifying Service Set Identifiers (SSIDs) and locations indicative of industrial networks, we gain empirical insight into Wi-Fi's penetration within the industrial domain. Crowdsourced analyses like this are widely recognized and employed in security research, offering scalable and empirical perspectives that are often difficult to obtain through traditional fieldwork~\cite{meas:TLS,mladenovall}. Specifically for Wi-Fi, similar methodologies have been successfully applied to assess WLAN abundance and security. For instance, the work in~\cite{LINDROOS2021108359} presented a systematic approach for WLAN assessment, validating WiGLE's utility as a comparative baseline. Similarly, the authors in~\cite{A:2022:Etta:WardrivingTechnique} utilized WiGLE-formatted data for wardriving-based security assessments, extracting statistics on encryption types, SSIDs, and vendor distributions. Furthermore, the work in~\cite{IN:2021:SecurityTrends:WiFi:Networks} examined global Wi-Fi security trends by incorporating multiple public datasets, including WiGLE. 

\noindent\textbf{\textit{Contribution:}} This work provides a timely and multifaceted contribution to the evolving landscape of industrial wireless security, spanning four key areas. We compile and present the first publicly available dataset~\cite{dataset} of 1,087 high-confidence industrial Wi-Fi networks globally. We offer comprehensive statistics on SSID naming conventions, security practices, vendor prevalence, and geospatial distribution within these industrial networks. We highlight and discuss the critical security risks associated with Wi-Fi integration in operationally sensitive industrial environments. We outline actionable best practices designed to enhance the security posture of wireless-enabled industrial systems. Through these contributions, our research establishes a baseline for understanding real-world Wi-Fi integration into industrial infrastructure and aims to stimulate essential future research into its security implications.

The rest of this paper is organized as follows. Section~\ref{S:data:coll} details the data collection framework. Section~\ref{S:analysis} analyzes the compiled dataset. Section~\ref{S:security:risks} discusses the security implications of Wi-Fi integration in industrial settings. Section~\ref{S:recom} outlines best practices for strengthening the security posture of wireless-enabled industrial systems. Section~\ref{S:limitations} summarizes the study's limitations. The last section concludes and identifies potential avenues for future research.

\section{Data Collection}
\label{S:data:coll}

\begin{figure}[!ht]
    \centering
\includegraphics[width=0.9\linewidth]{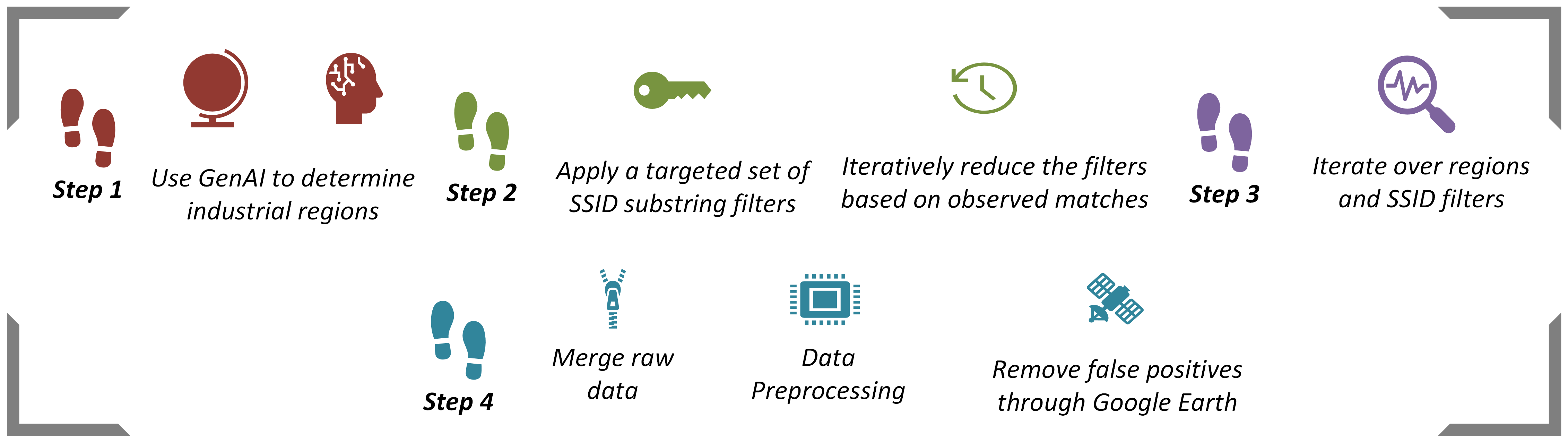}
    \caption{Workflow for identifying industrial wireless networks.}
    \label{F:workflow}
\end{figure}

To empirically assess global industrial Wi-Fi deployments, we designed and implemented a systematic data collection framework utilizing the WiGLE API~\cite{wigle}, as illustrated in Figure~\ref{F:workflow}.

\begin{figure}[!ht]
    \centering
\includegraphics[width=0.7\linewidth]{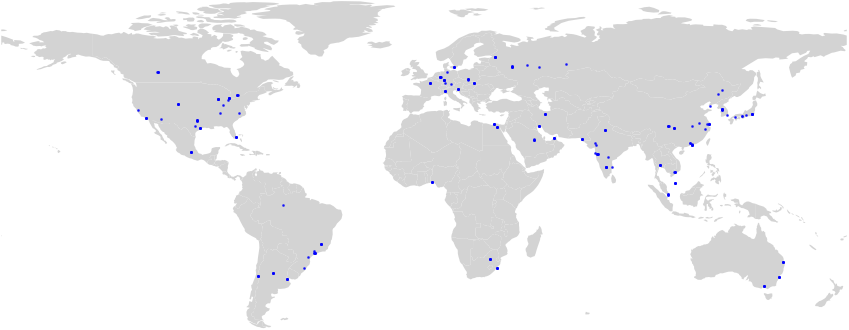}
    \caption{Search regions. The scanned locations cover a geographical area totaling $\approx75.000 km^2$.}
    \label{F:map}
\end{figure}

\noindent \textbf{\textit{Step 1 (Region selection):}} Our initial step involved compiling a list of 264 industrially significant regions worldwide. Each region is precisely defined by a geographic bounding box, using latitude and longitude coordinates, as depicted in Figure~\ref{F:map}. Region selection was facilitated by an AI-assisted process utilizing a Large Language Model (LLM), specifically ChatGPT Plus. This LLM leveraged its advanced analytical capabilities to synthesize insights from open-source intelligence, infrastructure datasets, and industrial geography literature, enabling the identification of global hotspots of industrial activity. Our selection criteria prioritized areas with significant manufacturing zones, logistics hubs, industrial ports, and energy infrastructure. To validate the LLM-generated list, we cross-referenced a sample of the selected regions with OpenStreetMap (OSM) and Environmental Systems Research Institute (ESRI) land-cover references~\cite{Yoo2025} and industrial cluster maps~\cite{ind:map}. Each selected region was then encoded into a JSON registry, detailing its \textit{code\_name}, geographic bounding box (\textit{lat\_min}, \textit{lat\_max}, \textit{lon\_min}, \textit{lon\_max}), and a brief justification for its inclusion. The specific prompt used to generate these regions is provided in Prompt~\ref{fig:region:prompt}.

\begin{figure}[ht]
\centering
\begin{tcolorbox}
\textit{Generate a JSON file containing geographic bounding boxes for approximately 200--300 industrially significant cities or regions around the world. Each entry should include the region’s name (\textit{code\_name}), latitude and longitude bounds (\textit{lat\_min}, \textit{lat\_max}, \textit{lon\_min}, \textit{lon\_max}), and a brief explanation (\textit{reason}) of its industrial relevance. The regions should cover diverse countries and industries, including manufacturing, electronics, automotive, and logistics hubs.}
\end{tcolorbox}
\caption{Prompt for generating industrial regions.}
\label{fig:region:prompt}
\end{figure}

\noindent \textbf{\textit{Step 2 (SSID Filtering):}} Subsequently, we employed a targeted set of 24 SSID substring filters to increase the likelihood of identifying industrial wireless networks from the initial regions dataset, acquired in the first step. These filters were meticulously designed to match SSIDs commonly associated with ICS, OT infrastructure, and vendor-specific deployments. Specifically, the list included manufacturer names such as \textit{Siemens}, \textit{Rockwell}, and \textit{Yokogawa}; control system terminology such as \textit{PLC}, \textit{SCADA}, \textit{HMI}, and \textit{Historian}; and infrastructure-related keywords like \textit{substation}, \textit{power grid}, and \textit{pump station}. The full list of SSID filters is presented in~\ref{tab:ssid:filt}. The SSID filters were iteratively refined based on observed matches during the region scanning process. This refinement allowed us to prioritize high-yield filters and eliminate ambiguous or low-signal terms, thereby improving both the precision and efficiency of the data collection pipeline. For instance, filters such as \textit{OPC Server}, \textit{Historian}, and \textit{General Electric} were eliminated early in the scanning phase due to consistently low or zero match rates across regions. Other filters, like \textit{site} and \textit{HMI} initially yielded high match counts but were discarded at a later stage after closer manual inspection of intermediate results. This inspection revealed their frequent use in non-industrial contexts, significantly reducing their specificity and relevance to the industrial domain.

\begin{table}[!htbp]
\centering
\caption{SSID substring filters used to identify industrial wireless networks. The character `\%` is used as a wildcard. Filters marked with \textsuperscript{(\dag)} were removed due to near-zero match rates, while those marked with \textsuperscript{(*)} were removed due to a predominant proportion of irrelevant matches.}
\label{tab:ssid:filt}
\resizebox{\textwidth}{!}{%
\begin{tabular}{|>{\raggedright\arraybackslash}p{4.5cm}|>{\raggedright\arraybackslash}p{4.5cm}|>{\raggedright\arraybackslash}p{4.5cm}|}
\hline
\textbf{ICS/OT Terminology} & \textbf{Infrastructure} & \textbf{Vendors} \\
\hline
\%PLC\%, \%SCADA\%, \%process\%control\%\textsuperscript{(\dag)}, \%control\%system\%, \%OPC\%server\%\textsuperscript{(\dag)}, \%Historian\%\textsuperscript{(\dag)}, \%Workstation\%, \%HMI\%\textsuperscript{(*)} &
\%substation\%, \%power\%grid\%, \%pump\%station\%\textsuperscript{(\dag)}, \%power\%station\%\textsuperscript{(\dag)}, \%waste\%water\%\textsuperscript{(\dag)}, \%site\%\textsuperscript{(*)}, \%plant\%, \%PCU\%\textsuperscript{(*)} &
\%Honeywell\%, \%Schneider\%, \%Rockwell\%\textsuperscript{(\dag)}, \%Emerson\%Electric\%\textsuperscript{(\dag)}, \%General\%Electric\%\textsuperscript{(\dag)}, \%Fortinet\%, \%Yokogawa\%, \%Siemens\% \\
\hline
\end{tabular}}
\end{table}

\noindent \textbf{\textit{Step 3 (API querying):}} To adhere to WiGLE's daily API request limit (100 requests per day), our collection script iterated through each region-filter pair, paginating results using the API's \textit{searchAfter} parameter. We checkpointed progress with a JSON state file, allowing the resumption of data collection across multiple days. Each successful query returned up to 100 networks per page, which were then stored in structured CSV files, organized by region and filter. Specifically, algorithm~\ref{alg:search} presents the approach for collecting Wi-Fi network data using the WiGLE API. The input consists of the list of predefined geographic regions $\mathcal{R}$, as identified in Step 1; a set of SSID-based heuristic filters $\mathcal{F}$ representing industrial naming patterns, as pinpointed in Step 2; and a daily query limit $Q$ imposed by the API. The algorithm begins by initializing the query counter and resuming from previously saved progress, represented by region index $i$, filter index $j$, and the pagination token $s$; in case the query results in more than 100 entries. For each selected region $r_i$ and filter $f_j$, the algorithm performs paginated API queries until no more results are available or the query budget $Q$ is exhausted. Each query returns a batch $\mathcal{B}$ of network metadata, which is appended to an output dataset $\mathcal{D}$. Once all pages for a given region-filter combination are retrieved, the corresponding results $\mathcal{D}_{i,j}$ are saved to a CSV file.

\resizebox{0.8\linewidth}{!}{%
\begin{minipage}{\linewidth}
\begin{algorithm}[H]
\caption{Industrial Wi-Fi Network Collection via WiGLE API}
\label{alg:search}
\KwIn{List of regions $\mathcal{R} = \{r_1, r_2, \dots, r_n\}$, SSID filters $\mathcal{F} = \{f_1, f_2, \dots, f_m\}$, max queries $Q$}
\KwOut{CSV files $\mathcal{D}$ containing network scan results}

Initialize query counter $q \leftarrow 0$\;
Load progress indices $i \in [1, n]$, $j \in [1, m]$, and search token $s$\;

\While{$q < Q$ and $i \leq n$}{
  Let $r_i$ be the current region and $f_j$ be the current filter\;
  
  \Repeat{search token $s = \emptyset$ \textbf{or} $q \geq Q$}{
    Send API query with $(r_i, f_j, s)$\;
    Receive results $\mathcal{B}$ and update $s$\;
    Append $\mathcal{B}$ to dataset $\mathcal{D}$\;
    $q \leftarrow q + 1$\;
  }

  Save $\mathcal{D}_{i,j}$ to file\;
}
\end{algorithm}
\end{minipage}%
}

\vspace{0.1cm}

\noindent \textbf{\textit{Step 4 (Data preprocessing):}} The raw scan results were initially stored hierarchically, with each geographic region in a dedicated directory and each SSID filter yielding a separate CSV file. Once the scanning phase was complete, all CSV files were merged into a unified dataset, resulting in 43,188 candidate industrial wireless networks, as detailed in Table~\ref{tab:dataset_reduction}. We then performed an initial data-cleaning process. This involved deduplication of identical entries, removal of false positives based on keyword heuristics (e.g., \textit{guest}, \textit{home}, \textit{cafe}), exclusion of records with invalid or empty location or SSID fields, filtering out networks with last visited timestamps before 2020, and elimination of spatial near-duplicates using a 1 km radius threshold via Haversine distance, which is used for determining distances between points on Earth's surface. Regarding the latter, we acknowledge that distinct WiFi networks with identical SSIDs may exist within 1km in dense industrial areas, leading to false negatives; this threshold was selected to strike a balance between redundant detections and data granularity.

\begin{table}[!ht]
\centering
\caption{Dataset reduction process through preprocessing and manual validation. $\Delta$ indicates the number of entries removed at the respective step.}
\label{tab:dataset_reduction}
\resizebox{0.7\textwidth}{!}{%
\begin{tabular}{|l|c|c|cc|}
\hline
\multirow{2}{*}{\textbf{Stage}} & 
\multirow{2}{*}{\textbf{Entries}} & 
\multirow{2}{*}{\textbf{$\Delta$ Entries}} & 
\multicolumn{2}{c|}{\textbf{Retention Rate}} \\
\cline{4-5}
& & & \textbf{Step (\%)} & \textbf{Total (\%)} \\
\hline
Raw merged dataset & 43,188 & -- & -- & -- \\
Preprocessed dataset & 11,973 & 31,215 & 27.7\% & 27.7\%  \\
Final dataset & 1,087 & 10,886 & 9.1\% & 2.6\%  \\
\hline
\end{tabular}}
\end{table}

Observe from Table~\ref{tab:dataset_reduction} that this preprocessing step reduced the initial data to 11,973 entries. To further eliminate false positives, we then manually verified the coordinates of each remaining entry using Google Earth. This enabled manual visual verification of the physical context of each wireless network, excluding those located in residential, commercial, or otherwise non-industrial environments. Only networks situated within visible industrial zones, such as factories, substations, energy facilities, or logistics hubs, were retained in the final dataset~\cite{dataset}. Our refined dataset ultimately contained 1,087 high-confidence industrial wireless networks, as also detailed in Table~\ref{tab:dataset_reduction}. The substantial reduction from the raw dataset can be attributed to multiple factors. Firstly, the raw data contained a large number of duplicate records, likely due to repeated scanning activities, overlapping coverage, or multiple users registering the same Wi-Fi network in the WiGLE database. Secondly, numerous entries were false positives retrieved by the SSID filtering heuristics. Thirdly, several entries lacked valid or complete information, including missing SSIDs, invalid geolocation coordinates, or outdated timestamps predating our threshold year. Finally, during manual verification through Google Earth, a significant number of entries were identified as being situated in residential, commercial, or non-industrial areas.

Nevertheless, the fact that 9.1\% of the preprocessed data represent high-confidence industrial networks suggests that systems commonly associated with OT environments are now more frequently exposed to external connectivity. Although this proportion may appear negligible, it is crucial to recognize that our dataset represents only a limited snapshot of the global industrial landscape; the scanned locations cover a constrained geographical area totaling $\approx75.000 km^2$. Additionally, the heuristic filters employed for SSID selection have inherent limitations, as they may fail to capture networks whose naming conventions do not align with predefined industrial keywords or are expressed in native languages, thus obscuring their identification. Furthermore, a substantial number of additional industrial wireless networks probably exist that were not detected. These might not be registered in public databases such as WiGLE, could be intentionally named using ambiguous or misleading SSIDs to conceal their presence and evade detection, or may be hidden due to private configurations, such as SSID cloaking, a practice that constitutes security by obscurity and should generally be avoided for robust protection. Consequently, the actual extent of wireless network proliferation in industrial settings is likely to be significantly greater than suggested by the available data~\cite{Li2017,VK:2023}.

\section{Analysis}
\label{S:analysis}

We analyzed 1,087 industrial Wi-Fi networks from our dataset~\cite{dataset} to assess their current state. Specifically, we measured the frequency of SSID filters to identify which industrial terms most consistently indicated a relevant network. We also analyzed the distribution of security protocols to evaluate the security practices employed in industrial Wi-Fi deployments and identified the most common hardware vendors by analyzing MAC address prefixes. Finally, we mapped country-level retention rates to reveal regional patterns in industrial Wi-Fi visibility.

\subsection{SSID naming conventions}

\begin{figure}[!ht]
    \centering
\includegraphics[width=0.5\linewidth]{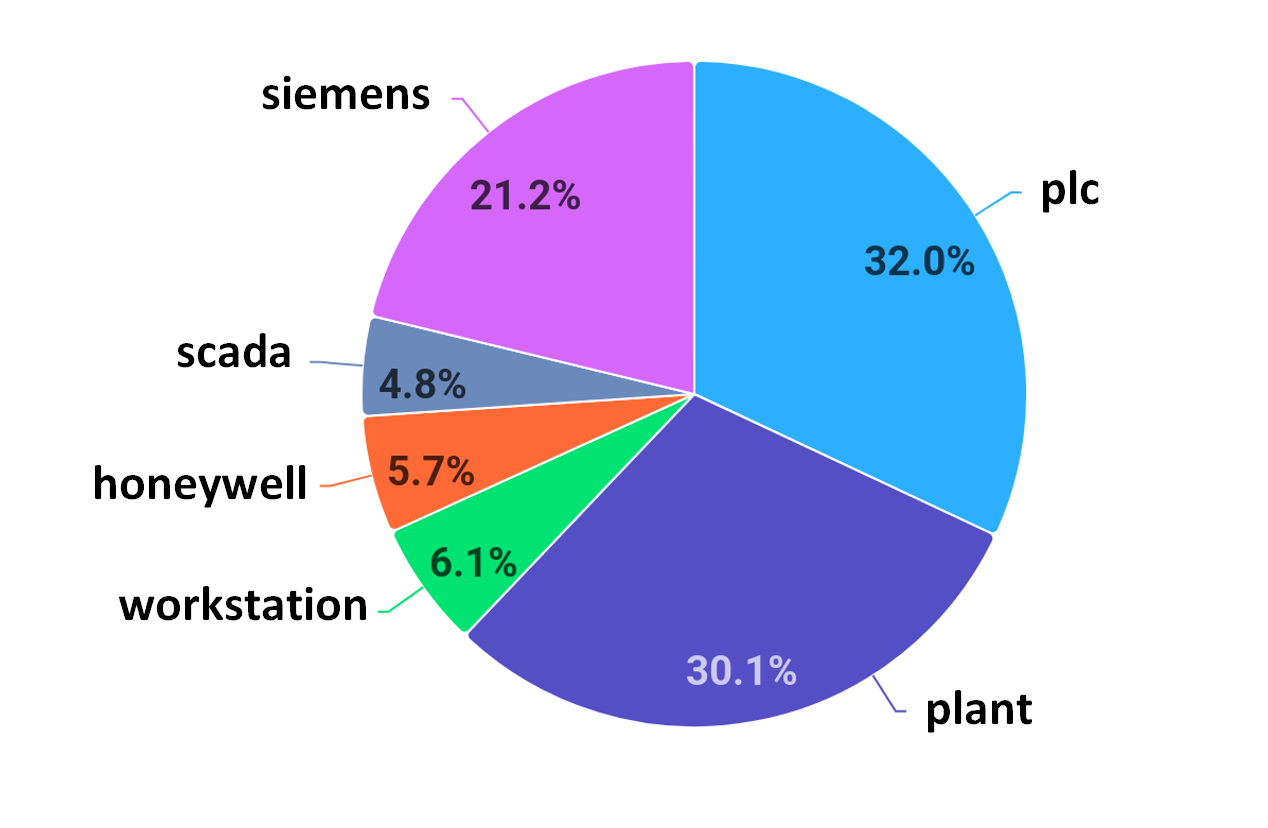}
    \caption{Keyword distribution in the dataset. Keywords representing less than 3\% are excluded.}
    \label{F:keywords:per}
\end{figure}

Figure~\ref{F:keywords:per} illustrates the most frequent SSID filters identified in the dataset, revealing common naming conventions in industrial Wi-Fi. PLC leads with 32.0\%, followed closely by \textit{plant} at 30.1\%. Other notable keywords are \textit{Siemens} (21.2\%), \textit{Workstation} (6.1\%), \textit{Honeywell} (5.7\%), and \textit{SCADA} (4.8\%). Keywords with smaller contributions are omitted for improved readability. The prevalence of these terms shows that industrial SSIDs often explicitly refer to OT (e.g., \textit{PLC}, \textit{SCADA}), infrastructure (e.g., plant, workstation), or vendors (e.g., \textit{Siemens}, \textit{Honeywell}).

\begin{figure}[htbp]
    \centering
    \begin{subfigure}{0.4\textwidth}
        \centering
        \includegraphics[width=5cm,height=2.7cm]{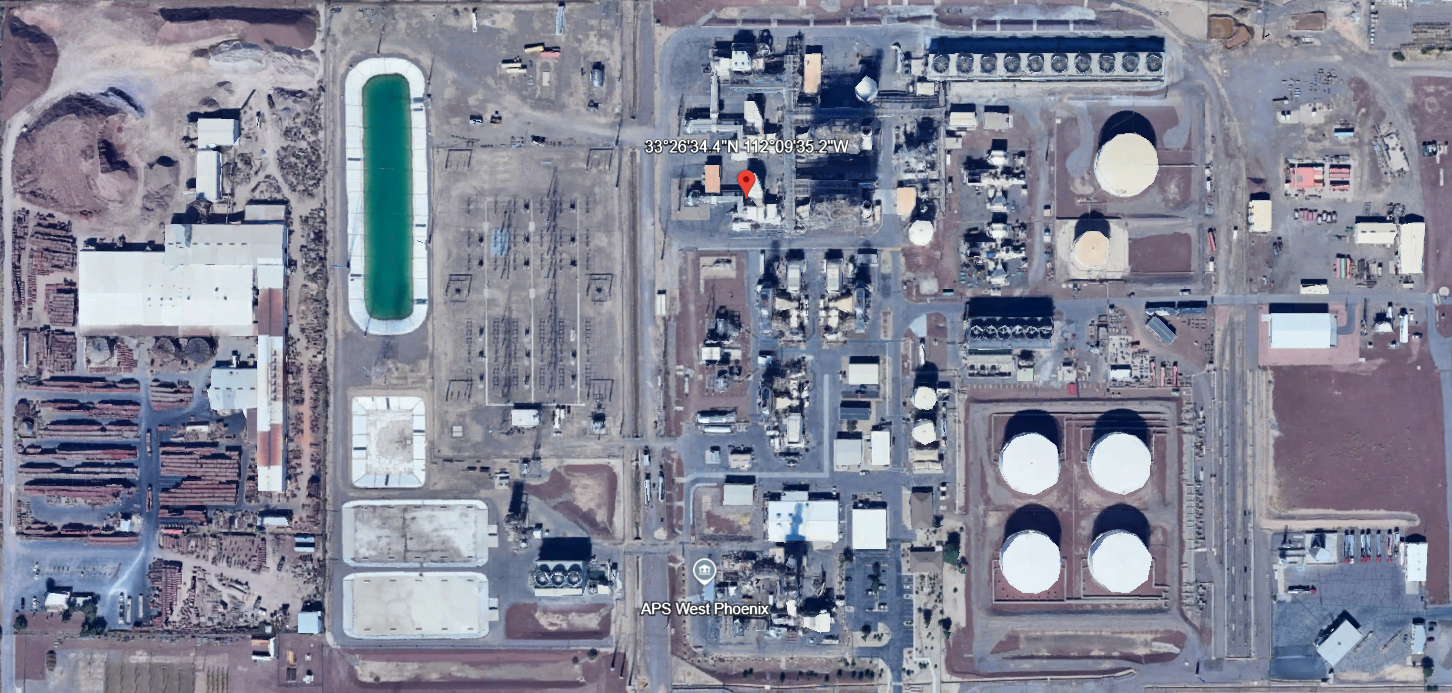}
        \caption{\%Siemens\%}
        \label{subf:screen1}
    \end{subfigure}
    \hfill
    \begin{subfigure}{0.4\textwidth}
        \centering
        \includegraphics[width=5cm,height=2.7cm]{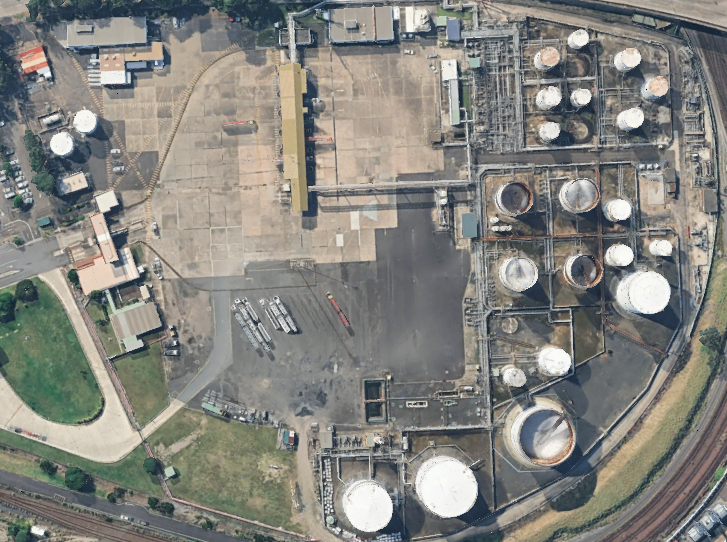}
        \caption{\%SCADA\%}
    \end{subfigure}

    \vspace{0.5cm}

    \begin{subfigure}{0.4\textwidth}
        \centering
        \includegraphics[width=5cm,height=2.7cm]{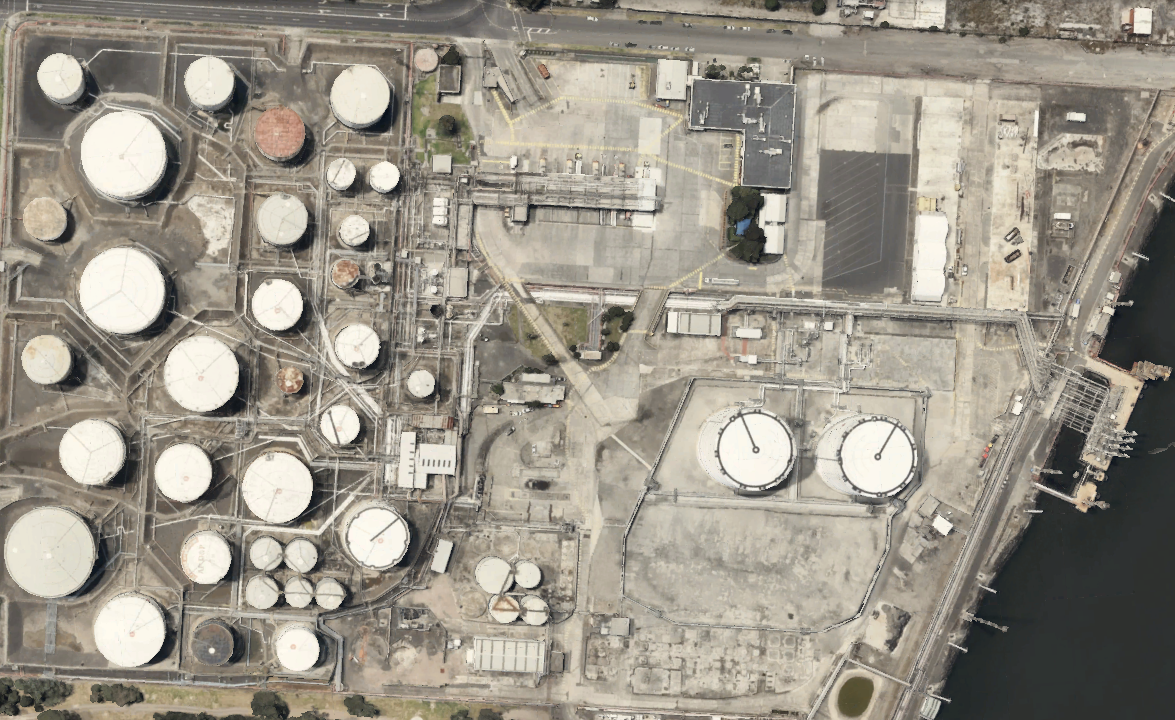}
        \caption{\%PLC\%}
    \end{subfigure}
    \hfill
    \begin{subfigure}{0.4\textwidth}
        \centering
        \includegraphics[width=5cm,height=2.7cm]{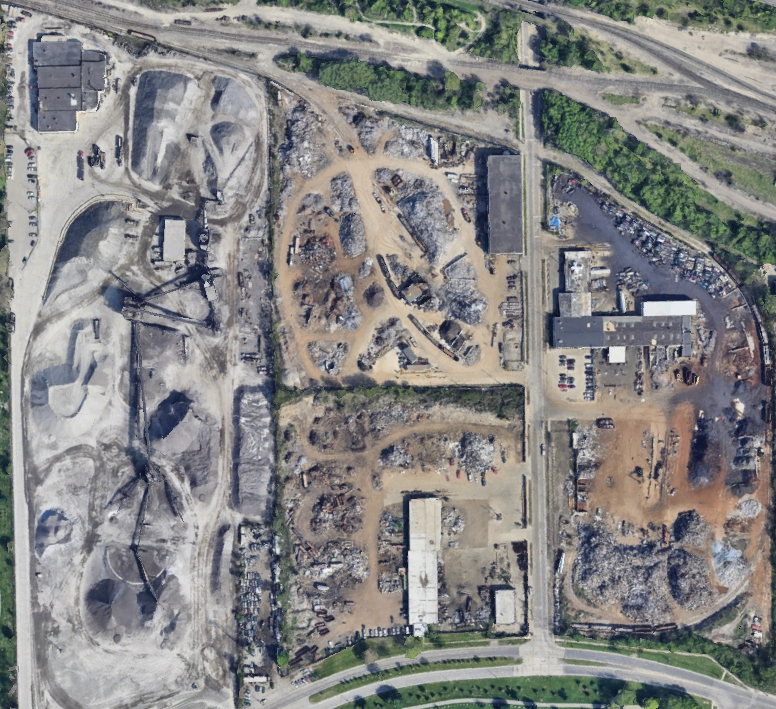}
        \caption{\%Plant\%}
        \label{subf:screen4}
    \end{subfigure}
    \caption{Screenshots of detected Wi-Fi networks associated with industrial systems.}
    \label{fig:screens}
\end{figure}

Figure~\ref{fig:screens} shows satellite imagery of industrial sites where Wi-Fi networks with SSIDs containing keywords like \textit{Siemens}, \textit{SCADA}, \textit{PLC}, and \textit{Plant} were detected. These are not generic or low-sensitivity environments; each location in subfigures~\ref{subf:screen1} to~\ref{subf:screen4} represents a Critical Infrastructure (CI) site. The clear presence of identifiable Wi-Fi infrastructure at these sites indicates that wireless technologies have indeed expanded into sectors traditionally known for tightly controlled networks and securely segregated OT ecosystems. While this direct labeling might be useful for internal management and system identification, it may also suggest a lack of security awareness. For internal operations, descriptive SSIDs can indeed simplify network administration, troubleshooting, and asset tracking. They help personnel quickly identify and connect to the correct network segment, especially in complex industrial facilities with numerous devices and varying access needs. This can improve operational efficiency and reduce human error. However, this convenience comes at a severe security cost. Such naming conventions enable heuristic filtering and facilitate a straightforward classification of industrially-relevant modules, presenting a clear information leakage vector, and exposing system roles and technologies to passive observers. In essence, these SSIDs are broadcasting valuable reconnaissance information to anyone within wireless range, whether they are a casual passerby or a dedicated malicious actor. The same methodology followed in this study can be weaponized by adversaries to quickly identify and target high-value ICS or OT environments. Furthermore, this specific approach to SSID naming highlights a broader issue: the absence of standardized security guidelines addressing SSID anonymization or the avoidance of disclosing critical system components, roles, or vendor identities.

\subsection{Security protocols}
\label{SS:encry}

\begin{figure}[!ht]
    \centering
\includegraphics[width=0.5\linewidth]{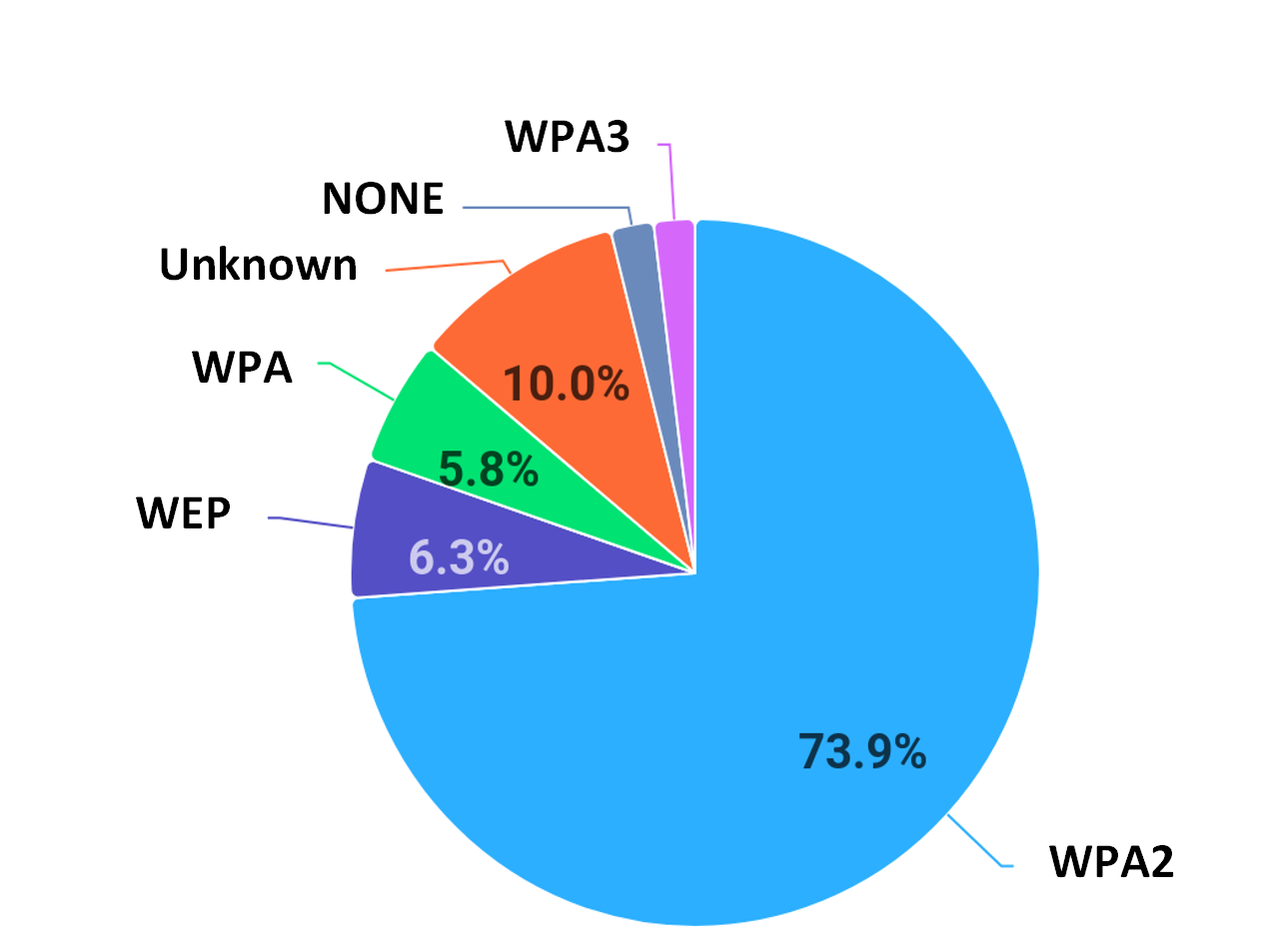}
    \caption{Distribution of Wi-Fi security types in the dataset. Types with less than 3\% are not labeled.}
    \label{fig:encryption}
\end{figure}

Figure~\ref{fig:encryption} illustrates the distribution of security protocols employed by the identified industrial Wi-Fi networks. The vast majority (73.9\%) utilize WPA2, while a non-negligible portion still relies on obsolete, less secure protocols: WEP (6.3\%) and WPA (5.8\%). Interestingly, for 10\% of the identified Wi-Fi networks, the security protocol was not registered. This could be attributed to inherent limitations of passive scanning or occasional gaps in metadata collection. Further investigation revealed that all networks identified within Russian territories were reported with an unknown security protocol. This strongly suggests that WiGLE may intentionally suppress encryption information for certain regions, possibly due to security concerns or adherence to local data policies. Finally, the most modern Wi-Fi security certification, WPA3, accounts for a mere 1.9\% of deployments, and networks with no security represent 2.0\% of the dataset.

Outdated Wi-Fi encryption protocols present serious security vulnerabilities, especially when used in CI environments. For example, the obsolete WEP is highly insecure, relying on short, repeating initialization vectors and a critically flawed implementation of the RC4 cipher~\cite{RC4:patterson}. This makes its keys extremely vulnerable and easy to crack, often in a matter of seconds with readily available tools. Similarly, WPA also has known weaknesses. While the 4-way handshake itself is openly exchanged during connection, Pre-Shared Key (WPA-PSK) networks are susceptible to offline dictionary/brute-force attacks if a weak passphrase is used. An attacker can capture this handshake and then attempt to crack the passphrase offline without needing further interaction with the network, effectively gaining access. A critical flaw in WPA-PSK is its lack of forward secrecy: if the pre-shared key is ever compromised, all past recorded traffic can be decrypted, as session keys are directly derived from this static key. Beyond passphrase cracking, the Temporal Key Integrity Protocol (TKIP) protocol has also been found partially vulnerable to cryptographic attacks. These vulnerabilities allow attackers to decrypt packets by exploiting weaknesses in TKIP's Message Integrity Code (MIC) and key scheduling, recovering portions of encrypted traffic which, while differing from a full key recovery, still allows for significant information disclosure~\cite{WEP:WPA:attacks,WPA:TKIP:vanhoef}. Furthermore, attackers can inject forged frames by crafting and injecting malicious data packets into a WPA/TKIP network, potentially leading to Denial-of-Service (DoS) or other undesirable actions, even without fully recovering the encryption key. Practical attacks against TKIP have also demonstrated the ability to recover the Michael message authentication key in minutes, enabling further exploitation.

Even WPA2 has been proven vulnerable under certain conditions. A prominent example is the Key Reinstallation Attacks (KRACK)~\cite{KRACK:vanhoef,KRACK1:vanhoef}, which targeted a protocol-level design flaw in the WPA2 4-way handshake. Another notable vulnerability is Kr00k (CVE-2019-15126), which revealed that many Wi-Fi devices implementing WPA2 (and WPA) could transmit certain data frames unencrypted under specific circumstances, such as after disassociation or deauthentication. Furthermore, WPA2-PSK networks, much like their WPA counterparts, remain susceptible to offline dictionary/brute-force attacks if a weak passphrase is used, as the 4-way handshake can be captured and cracked offline. Critically, WPA2-PSK also lacks forward secrecy, meaning that if the pre-shared key is ever compromised, all past recorded encrypted traffic can be decrypted. Moreover, the study in~\cite{awid} provides a real-world attack dataset against WPA2-enabled Wi-Fi networks. This dataset highlights vulnerabilities such as the absence of Protected Management Frames (PMF) (an optional feature in WPA2, mandated in WPA3), which enables deauthentication and disassociation attacks leading to DoS. The same work also demonstrates authentication flooding and fake access point injections, often facilitating Man-in-the-Middle (MitM) scenarios. Additionally, it reveals how MAC spoofing and replay attacks exploit insufficient frame protection, further highlighting how active adversaries can interfere with Wi-Fi operations, even without directly breaking cryptographic keys. Altogether, relying on outdated or misconfigured Wi-Fi security (WEP, WPA, or even unpatched WPA2) leaves networks open to well-documented attacks. These range from classic WEP key-cracking tools to sophisticated handshake manipulations, all of which can lead to unauthorized network access and traffic manipulation by a variety of threat actors.

In this context, WPA3 was designed to address many of the inherent flaws of its predecessors. Specifically, for WPA3-Personal, typically used in Small Office/Home Office (SOHO) environments, a key improvement is the Simultaneous Authentication of Equals (SAE), an authentication and key exchange method derived from the Dragonfly Key Exchange (defined in RFC 7664) and incorporated into the IEEE 802.11-2016 and onwards standards. WPA3-Enterprise deployments continue to utilize 802.1X/EAP for authentication. Notably, SAE is resistant to offline dictionary attacks and provides forward secrecy. WPA3 also imposes frame protection through the mandatory use of PMF, introduced with the 802.11w amendment, and mandates the use of modern cryptographic algorithms. However, despite its security enhancements, WPA3 is not without weaknesses. Recent studies have uncovered several critical vulnerabilities in both its design and Access Point (AP) firmware implementations. Importantly, early analyses of SAE revealed protocol-level weaknesses~\cite{vanhoef:dragonblood}, including DoS via resource exhaustion, and downgrade attacks in WPA3's transitional mode, forcing fallback to WPA2. These analyses also exposed side-channel leaks enabling offline password cracking. Subsequent studies demonstrated that WPA3 networks remain vulnerable to various DoS vectors. For instance, an adversary can overload APs by exploiting the CPU-intensive SAE exchange~\cite{Chatz:wpa3:dos}, or disrupt communications by injecting spoofed unprotected control frames, such as \textit{BlockAck} frames~\cite{Bl0ck}. Implementation flaws further compound these issues. The work in~\cite{wpaxfuzz}, for example, exposed numerous bugs in management frame handling code that could be exploited to crash or destabilize devices through fuzzing, even in the presence of WPA3's mandatory PMF mechanism~\cite{awid3}. Even WPA3's newer features have shown vulnerabilities. The SAE-PK mechanism, for instance, can leak its private key if a weak random number generator is used and is susceptible to time–memory trade-off attacks (rainbow tables), allowing efficient offline recovery of the password~\cite{vanhoef:WPA3-PK}. Finally, fundamental design oversights in Wi-Fi's frame queuing were found to bypass WPA3 encryption entirely. Specifically, an attacker can trick an AP into sending buffered frames encrypted with an all-zero (null) key or plaintext, and can also force client disconnections via manipulated queue control flags~\cite{schepers2023framing}.

\subsection{Vendor landscape}

\begin{figure}[!ht]
    \centering
\includegraphics[width=0.8\linewidth]{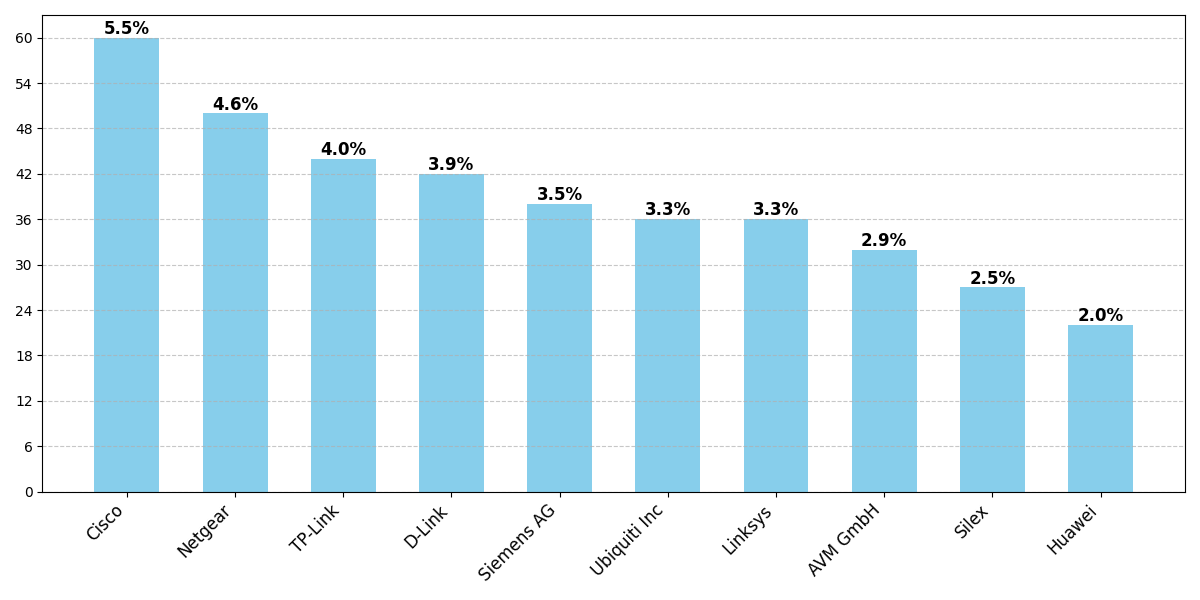}
    \caption{Top 10 AP manufacturers in the dataset.}
    \label{F:manu:top10}
\end{figure}

Analysis of Wi-Fi AP manufacturers revealed prominent vendors, identified by matching the MAC address prefixes (OUIs) of the detected networks with a MAC vendor lookup database~\cite{macvendor}. As illustrated in Figure~\ref{F:manu:top10}, well-known vendors such as Cisco (5.5\%), Netgear (4.6\%), and TP-Link (4.0\%) appear most frequently in the dataset, indicating a significant footprint in industrial and enterprise wireless infrastructure. Notably, the presence of Siemens AG (3.5\%) among the top manufacturers further confirms the dataset's industrial relevance, given Siemens' strong association with automation and control systems. Vendors such as D-Link, Ubiquiti, and Linksys also contribute sizable shares. Interestingly, 24.3\% of the identified AP devices are associated with unknown vendors. This suggests that their manufacturers may not be registered in the MAC vendor lookup API, or that the address blocks used are either anonymized, unassigned, or derived from obscure or non-standard hardware sources.

Interestingly, SOHO vendors like Huawei (2.0\%), Xiaomi, and Zyxel are underrepresented in the dataset, suggesting that commodity-grade hardware is -- as anticipated -- less common in industrial contexts. In contrast, although vendors like Netgear, TP-Link, and D-Link are widely recognized for their SOHO product lines, they also offer a broad range of enterprise-grade APs. This trend reflects a clear preference for more robust, managed, and security-capable hardware in critical environments. Enterprise APs offer crucial advantages over SOHO ones, including centralized management, stronger security features like WPA3-Enterprise, 802.1X authentication, and certificate-based access control, higher reliability under varying workloads, enhanced environmental durability, and long-term firmware support. These features are essential for maintaining security and operational continuity in complex industrial environments. Despite their general scarcity, the presence of SOHO vendors in the dataset suggests that these devices may be frequently repurposed for industrial environments. This could reflect cost-driven decisions, uninformed legacy deployments, or overlay network expansions implemented without centralized oversight. In many cases, such deployments offer short-term convenience but typically lack the hardened security features and lifecycle guarantees of enterprise solutions. This potentially introduces vulnerabilities and increases the risk of misconfiguration, outdated firmware, or insufficient access control, all of which are critical concerns in CI settings.

\subsection{Geospatial distribution}

\begin{figure}[!ht]
    \centering
\includegraphics[width=0.8\linewidth]{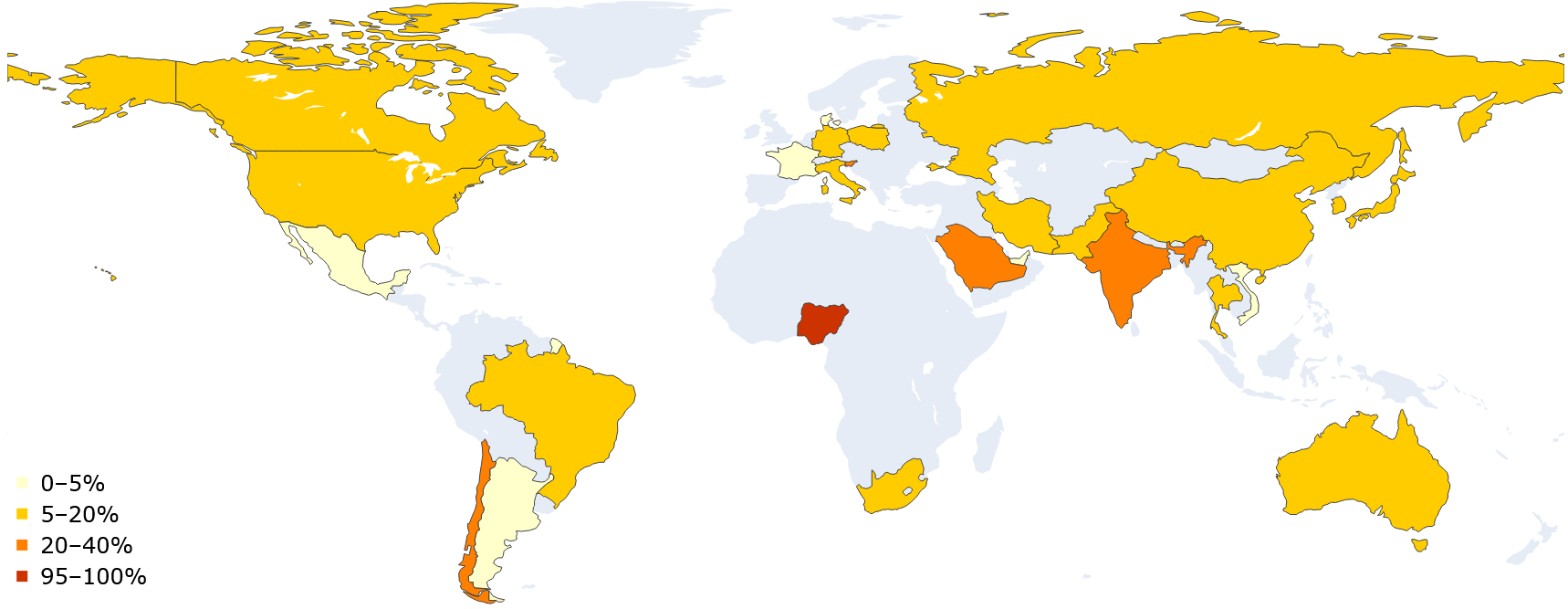}
    \caption{Heatmap of dataset distribution by country. Countries are shaded by the percentage of industrial Wi-Fi networks retained in the final dataset, relative to the preprocessed dataset.}
    \label{F:heatmap}
\end{figure}

Figure \ref{F:heatmap} illustrates the geospatial distribution of the dataset, with countries shaded based on the percentage of industrial Wi-Fi networks retained after the filtering steps outlined in Section~\ref{S:data:coll}. The white (unshaded) regions on the map were intentionally omitted from the search due to their relatively lower industrial relevance and WiGLE's daily API request restrictions, as discussed in the data collection process in Section~\ref{S:data:coll}. We opt to visualize the percentage of retained industrial networks relative to each region's preprocessed data, rather than relative to the absolute count; using global-normalized values would disanalogously highlight countries with larger scan volumes, potentially obscuring insights from smaller or less-scanned regions.

Among the analyzed countries, Nigeria (100\%), Saudi Arabia (33.3\%), Chile (23.4\%), and India (20.6\%) show some of the highest retention percentages, all exceeding 20\%. This suggests a strong match between their wireless industrial network footprints and the keyword-based filtering heuristics we used. Most of the regions searched, including the United States (9.3\%), Brazil (8.1\%), Australia (10.4\%), and major European and Asian industrial economies, fall within the 5–20\% range. These moderate retention levels might indicate a broader diversity in SSID naming or a larger proportion of non-industrial networks within the scanned areas. On a positive note, this trend could also suggest more cautious SSID labeling practices in these regions, discouraging the exposure of sensitive operational details in network identifiers. Finally, a subset of countries, including France (0.3\%), Vietnam (1.47\%), Argentina (1.97\%), Denmark (2.04\%), the United Arab Emirates (2.44\%), and Mexico (3.09\%), display retention percentages below 5\%. These results may point to less descriptive SSID conventions in industrial deployments, a lower density of industrial Wi-Fi infrastructure within the scanned regions, or more stringent anonymization practices. Additionally, the use of SSIDs in native languages or non-English scripts could hinder detection by keyword-based filtering heuristics. In some cases, like France, the near absence of identifiable industrial SSIDs may also reflect stricter adherence to operational security policies.

\section{Security implications}
\label{S:security:risks}

The integration of Wi-Fi into industrial systems significantly expands the attack surface, and the known wireless vulnerabilities are especially concerning in converged IT/OT environments like ICS and smart manufacturing. For example, in a scenario where factory engineers use Wi-Fi–enabled tablets for Human-Machine Interface/Supervisory Control and Data Acquisition (HMI/SCADA) access, an attacker can remotely crack a weakly protected industrial Wi-Fi network. This allows them to pivot directly into the control system and gain network access to field control devices such as Programmable Logic Controllers (PLC) and Remote Terminal Units (RTU). Even if facilities are well-isolated and segmented, e.g., following the well-known Purdue Enterprise Reference Architecture (PERA), insecure Wi-Fi can still introduce a backdoor. Specifically, corporate or contractor Wi-Fi networks that are bridged (even indirectly) to the plant network become attractive targets. Adversaries can compromise a poorly secured corporate WLAN, subsequently tunneling into the control system environment. Similarly, an initial compromise on the business logistics layer through a wireless link can enable malicious actors to move laterally~\cite{smilio} within the corporate network and eventually reach the lower layers of field control and devices.

Similarly, nearest-neighbor attacks involve exploiting devices or networks located in proximity to industrial sites. In settings like industrial parks, shared facilities, or urban sites, adjacent Wi‑Fi networks may expose attack paths. If these neighboring systems are dual-homed, a compromise can serve as a bridge into otherwise segmented environments. Even without direct connections, attackers may leverage signal leakage or misconfigured association behaviors to laterally traverse network boundaries and bypass perimeter defenses. Recall that this tactic was exemplified by the Russian APT28 group during cyber operations linked to the Ukraine invasion, where attackers breached internal networks by first compromising Wi‑Fi-connected systems in adjacent offices, exploiting proximity and poor isolation to evade defenses~\cite{volexity2024}.

This is exacerbated by the fact that many industrial protocols such as MODBUS, DNP3, and proprietary fieldbus protocols lack built-in encryption or authentication; they often send commands in cleartext and trust the network's integrity. Consequently, once an attacker is on the ICS network, they can intercept, spoof, or alter control commands and sensor readings with relative ease. For example, an attacker in a MitM scenario could inject false sensor data or send unauthorized control signals, potentially disrupting the physical process. This is because, unlike IT systems which often employ robust transport-layer security, many OT devices still have predictable sequence numbers and other weaknesses that make MitM attacks feasible, meaning that a Wi-Fi MitM can be leveraged to cause real process manipulation. Indeed, even passive eavesdropping on an ICS Wi-Fi network can leak sensitive operational data. For instance, decrypting wireless traffic might reveal sensor measurements or control setpoints that could enable an attacker to perform more damaging actions.

These concerns have not gone unnoticed by leading industrial vendors, many of whom have acknowledged that their products are vulnerable to Wi-Fi–based attacks~\cite{affected1,affected2}. For example, Cisco has confirmed that its 829 Industrial Integrated Services Routers and Industrial Wireless 3700 Series APs are susceptible to the KRACK vulnerability~\cite{KRACK:vanhoef}. Similarly, Rockwell Automation's Stratix 5100 APs, along with several Siemens product lines such as SCALANCE, SIMATIC, RUGGEDCOM, and SINAMICS, have been identified as affected. In this context, Cisco issued a technical report~\cite{cisco:ebook} emphasizing the security risks posed by wireless connectivity in ICS. It outlines how attackers can exploit weaknesses in wireless networks to gain unauthorized access to essential components such as SCADA systems and PLCs, potentially jeopardizing the integrity and reliability of industrial operations. In the same vein, Kaspersky Lab ICS CERT has extensively documented how vulnerabilities in wireless protocols can be exploited in attacks targeting industrial systems~\cite{kasp:notes}. Overall, the security implications of outdated Wi-Fi protections in OT environments extend significantly beyond passive threats like eavesdropping or data theft, as detailed in Section~\ref{SS:encry}. They introduce the concrete possibility of attackers triggering physical disruptions or safety incidents by exploiting wireless links in converged IT/OT networks. Therefore, safeguarding against these threats necessitates maintaining strong, up-to-date wireless configurations and eliminating deprecated protocols. Put differently, proactively securing wireless communication is a foundational requirement for preserving both operational continuity and the safety of physical processes.

\section{Best practices}
\label{S:recom}

This section outlines best practices that can be employed to reduce attack surfaces and promote the safe integration of wireless technologies into CI environments, drawing from well-known standards such as NIST SP 800-53, NIST SP 800-82, and IEC 62443.

\noindent \textbf{Descriptive SSIDs:} Industrial entities should refrain from using SSIDs that could reveal sensitive information like vendor names, component types, or facility identifiers. Instead, they should employ randomized or neutral SSIDs, following anonymization practices to reduce the risk of passive reconnaissance. This approach makes wardriving and other passive surveillance techniques less likely to yield actionable intelligence for adversaries. Such obfuscation aligns with the operational security (OPSEC) measures described in NIST SP 800-53, which aim to minimize information leakage and increase the effort required for network fingerprinting or targeted intrusion attempts.

\noindent \textbf{Centralized management and monitoring:} Wi-Fi deployments in industrial ecosystems should leverage enterprise-level APs, which typically offer centralized management capabilities. This enables continuous visibility into network operations, facilitates real-time monitoring, and ensures consistent enforcement of security policies across all connected devices. In such industrial settings, where availability is paramount, these features are critical for maintaining compliance, detecting unauthorized access attempts, and promptly isolating compromised network segments. Equally important, integration with Security Information and Event Management (SIEM) modules can further enhance situational awareness and incident response readiness.

\noindent \textbf{Network segmentation:} Industrial networks should be segmented to isolate OT systems from general IT traffic. This separation minimizes the risk of lateral movement attempts. Even when Wi-Fi connectivity is required in the OT segment it should be provisioned on dedicated, access-controlled Virtual Local Area Networks (VLANs) that enforce strict traffic isolation. Additionally, wirelessly accessible OT-specific segments should be subject to tailored firewall rules, role-based access controls, and continuous monitoring to ensure minimal exposure to the uppermost layers of the corporate network, aligning with security frameworks such as IEC 62443 and NIST SP 800-82, which emphasize defense-in-depth strategies in industrial environments.

\noindent \textbf{Audits and penetration testing:} Security assessments, including Wi-Fi penetration tests, should be periodically conducted to ensure deployed wireless infrastructure aligns with current security policies and industry standards. Penetration testing can uncover vulnerabilities that may not be evident through routine checks. Additionally, site surveys using spectrum analysis tools can detect unintended signal propagation beyond physical perimeters, mitigating risks such as eavesdropping or unauthorized device associations. These routine assessments also verify that device and firmware hygiene is maintained and deprecated configurations are removed.

\noindent \textbf{Personnel awareness:} Operators and administrators of industrial systems should undergo targeted training focused specifically on wireless security, with an emphasis on Wi-Fi-related threats. Beyond theoretical knowledge gained from tabletop exercises, organizations should adopt practical training methods, such as cyber ranges~\cite{KAMPOURAKIS2025103917} and hands-on labs, to simulate real-world attacks. This helps personnel develop the technical skills and situational awareness necessary to identify Wi-Fi-related vulnerabilities. By cultivating wireless-specific expertise, organizations enhance their frontline defenses and ensure that critical systems operators can confidently secure and manage Wi-Fi infrastructure in industrial environments.

\noindent \textbf{Physical exposure and signal leakage:} Physical security controls and radio containment techniques should be employed to restrict Wi-Fi signal propagation beyond the boundaries of industrial facilities. This includes the use of directional or low-gain antennas, strategic AP placement, and physical barriers. Limited signal leakage minimizes the risk of unauthorized access or passive reconnaissance by adversaries positioned near the facility. In addition, periodic wireless site surveys should be conducted to identify areas of excessive signal bleed or shadow zones, allowing for adjustments that optimize both security and performance.

\section{Limitations}
\label{S:limitations}

While the analysis presented in Section~\ref{S:analysis} offers meaningful insights into the state of industrial Wi-Fi deployments and the industry's transition towards IT-based infrastructure, several limitations must be acknowledged. First, the dataset was compiled through a selective and heuristic-based querying process via the WiGLE API, constrained by daily request limits and guided by industrial relevance. Naturally, this approach narrows the scope of coverage, excluding regions deemed less relevant and potentially overlooking industrial deployments that do not match the applied SSID heuristics. Second, the spatial distribution of the data is uneven. Even within countries included in the data collection process, the search was limited to specific cities or known industrial hubs rather than offering national coverage. This localized focus, while useful for targeting high-density industrial zones, limits the generalizability of observed trends and may bias results toward regions with higher visibility, scanning activity, or more extensively documented infrastructure. 

Third, the analysis relies heavily on SSID-based heuristics and MAC address prefixes for filtering and vendor classification, respectively. While these techniques are easy to implement, they can also be brittle. Specifically, industrial networks using non-descriptive SSIDs, native language labels, or obfuscated naming schemes may evade detection. Similarly, MAC prefixes from unregistered or ambiguous vendors may be misclassified or excluded. Finally, the collected data reflects only data that has been passively observed and uploaded to WiGLE. This introduces further bias stemming from who performs the data collection, when and where it occurs, what scanning devices are used, and which devices and networks are visible during scanning. Keeping these limitations in mind, the results should be interpreted as indicative of observable trends in industrial Wi-Fi deployment rather than as comprehensive or globally representative findings. Additionally, it is important to emphasize that the core contribution of this paper is not to provide exhaustive global coverage, but rather to demonstrate that Wi-Fi is indeed being integrated into industrial settings in the context of the ongoing IT/OT convergence. In other words, the observed trends and case studies offer empirical evidence that wireless technologies are increasingly present in operational environments traditionally characterized by wired and siloed architectures, even though the findings are shaped by the scope and constraints of the data collection process.

\section{Conclusions}

This work provides the first, to our knowledge, data-driven investigation into the real-world adoption of industrial Wi-Fi deployments using the widely recognized WiGLE database. To achieve this, we applied a thorough data collection and filtering process grounded in SSID keyword heuristics and manual validation, compiling a high-confidence dataset of 1,087 industrial wireless networks distributed across diverse regions. The results successfully substantiate the claim that wireless technologies, particularly Wi-Fi, are increasingly entering formerly air-gapped industrial environments, driven by the demands of Industry 4.0+ and the broader convergence of IT and OT infrastructures. Our analysis uncovered several key insights. First, a substantial number of industrial Wi-Fi networks use descriptive SSIDs that expose vendor names, operational roles, or infrastructure types. This facilitates potential passive reconnaissance and undermines security through information leakage. Second, while the majority of networks adopt WPA2, a concerning minority still rely on obsolete or weak security implementations associated with WEP and WPA. Third, the vendor landscape revealed a mix of enterprise and commodity-grade APs. Finally, the geospatial distribution of networks highlighted region-specific variations in SSID practices and detection visibility, possibly shaped by linguistic, policy, and deployment factors. Beyond empirical observations, this study underscores the broader security implications of deploying Wi-Fi in operationally sensitive environments. Wireless connectivity, while enhancing flexibility and efficiency, also introduces new attack vectors, including reconnaissance for network mapping and vulnerability identification, cyber-physical intrusion into control systems, lateral movement to compromise operational assets, and process disruption with potentially severe consequences for production, safety, and environmental control. These security concerns highlight the urgent need for standardized guidelines and best practices tailored to the unique profile of industrial wireless systems. Looking forward, the dataset and findings presented lay a foundation for future research, including the development of more robust and scalable scanning methodologies that go beyond SSID-based heuristics and the creation of a Wi-Fi-specialized cyber range tailored to industrial use cases.

\begin{credits}
\subsubsection{\ackname}This work is supported by the Research Council of Norway through the SFI Norwegian Centre for Cybersecurity in Critical Sectors (NORCICS) project no. 310105 and by the European Union through the Horizon 2020 project PERSEUS (Grant No. 101034240).
\end{credits}
%
%
%
 \bibliographystyle{splncs04}
 \bibliography{mybibliography}

\end{document}